\documentclass[twocolumn,epjc3]{svjour3}  
\usepackage[utf8]{inputenc}
\smartqed  % flush right qed marks, e.g. at end of proof
\RequirePackage{graphicx}
\RequirePackage{mathptmx}      % use Times fonts if available on your TeX system
\RequirePackage[colorlinks,citecolor=blue,urlcolor=blue,linkcolor=blue]{hyperref}
\usepackage{graphics}
\usepackage[shortlabels]{enumitem}
\usepackage{orcidlink}

\usepackage{amsmath}
\usepackage{amssymb}
\usepackage{pifont}
\usepackage[switch]{lineno}
\usepackage{cite}

\usepackage[normalem]{ulem}

\newcommand{\NLOAccess}{\texttt{NLO\-Ac\-cess} }
\newcommand{\HELACOnia}{\texttt{HE\-LAC\--Onia} }

\newcommand{\MG}{\texttt{MG5} }
\newcommand{\Flask}{\texttt{Flask }}

\DeclareMathAlphabet{\pazocal}{OMS}{zplm}{m}{n}
\newcommand{\Q}{\pazocal{Q}}

\journalname{Eur. Phys. J. A}
\begin{document}

\title{{\tt NLOAccess}: automated online computations for collider physics}

\titlerunning{{\tt NLOAccess}}        % if too long for running head

\author{Carlo Flore \orcidlink{0000-0002-1071-204X}\thanksref{e1,addr1} }

\thankstext{e1}{e-mail: carlo.flore@ijclab.in2p3.fr (corresponding author)}


\institute{Université Paris-Saclay, CNRS, IJCLab, 91405 Orsay, France \label{addr1}}

\date{Received: date / Accepted: date}

\maketitle

\begin{abstract}
We present {\tt NLOAccess}, an online platform acting as a virtual access to automated perturbative computations of physical ob\-serv\-a\-bles related to collider physics within collinear factorisation. We discuss the design of the project and illustrate how, with few steps and without any need of pre-code or code compilation, a user can generate events and calculate cross sections related for instance to quarkonium production and/or heavy-ion collisions at leading and next-to-leading order in $\alpha_s$ using {\tt Mad\-Graph5\_aMC\-@NLO} or {\tt HE\-LAC\--Onia}. Finally, we report about the obtained results in terms of number of users and runs, and we illustrate the future plans of the project. 

\end{abstract}

\section{Introduction}
\label{sec:intro}

Perturbative numerical computations are key ingredients in different research areas in high-energy physics for searches of new particles, for studies of the quark-gluon-plasma properties or of heavy-quark (onium) production. On the one hand, there are many available codes allowing for the computation of cross sections for different reactions. They enable their users to calculate such physical observables for several processes at higher orders in Quantum Electro- and Chromo-dynamics (QED and QCD) or within the full Standard Model (SM), that is when one considers all the couplings and vertices in the QCD+Electroweak theory. Some examples of such a kind of these codes can be found in Refs.~\cite{Mangano:2002ea,Wang:2004du,Gavin:2010az,Alioli:2010xd,Bevilacqua:2011xh,Hamilton:2012np,Campbell:2019dru}. On the other hand, not all of them behave in a fully automated fashion, {\it i.e.}~the generation of dedicated codes starts from a generic request, (for instance, a process in a certain model -- within or beyond the SM). Some examples of automated codes and event generators can be found in Refs.~\cite{Corcella:2000bw,Kilian:2007gr,Belyaev:2012qa,Sherpa:2019gpd}.
Noticeable exceptions are represented by two programs, {\tt Mad\-Graph5\_aMC\-@NLO}~\cite{Alwall:2014hca} (\MG in what follows) and \HELACOnia\cite{Shao:2012iz,Shao:2015vga}. They are fully automated codes, respectively at leading and next-to-leading order (LO and NLO) for SM and Beyond SM (BSM) calculation, and at LO and approximated higher orders for single and multiple quarkonium production within the SM.

Generally speaking, the full automation offers a vast flexibility in terms of usage of different BSM models (see for instance the large quantities of the so-called Universal FeynRules Output (UFO) models~\cite{Degrande:2011ua} for {\tt MG5\_aMC@NLO}\footnote{A non exhaustive list of UFO models can be found at \href{https://madgraph.mi.infn.it/models_db.dat}{https://madgraph.mi.infn.it/models\_db.dat}}) and of generation of various processes within a model. However, at the same time, it requires these codes to be as much user-friendly as possible. Another potential complication is represented by their compilation, linking and usage with external libraries/codes like for instance {\tt LHAPDF}~\cite{Buckley:2014ana} or {\tt PYTHIA}~\cite{Bierlich:2022pfr}. 

To avoid these potential drawbacks, online tools surely represent a valuable asset. They allow any non-expert to produce scientific output without needing any deep technical knowledge of the codes that are used and, in turn, they can be used to learn how to use such codes, to train non-experts and, potentially, to help in stimulating new ideas and produce new scientific results. One example of an online tool is for instance {\tt APFEL Web}~\cite{Carrazza:2014gfa}, that allows one to visualise the parton distribution functions (PDFs) and plot parton luminosities at the same time for different users. The project we present here, {\tt NLOAccess}, goes beyond the relatively simple visualisation of PDFs and computation of parton luminosities.

\NLOAccess is one of the two Virtual Accesses among the different Working Packages of the EU Horizon 2020 integrated initiative called ``STRONG-2020". As mentioned above, it gives access to automated tools allowing anyone to compute observables related to different scattering processes involving hadrons, without any need of writing down any part of code before a run. More information on \NLOAccess can be found at \href{https://nloaccess.in2p3.fr}{https://nloaccess.in2p3.fr}. In the following, we will also refer to the codes available on \NLOAccess as {\it tools}. At present, the two available tools on \NLOAccess are \MG and {\tt HELAC-Onia}. \NLOAccess and its tools are accessible at this address:
\begin{center}
\href{https://nloaccess.in2p3.fr/tools/}{https://nloaccess.in2p3.fr/tools/}
\end{center}
The paper is organised as follows: Section~\ref{sec:design} illustrates the design of \NLOAccess in terms of its server scheme, while in Section~\ref{sec:results} we explain how to obtain results from the online platform. Finally, in Section~\ref{sec:conclusions} we gather the conclusions and ideas for the future of such a project.  

\section{Design}
\label{sec:design}

Let us first present how the \NLOAccess platform is designed and illustrate the different sides needed to handle multiple computations for several users at the same time.

\NLOAccess relies on {\tt Flask}\footnote{\href{https://flask.palletsprojects.com}{https://flask.palletsprojects.com}}, a \texttt{Python} micro-framework, that allows one to design a web application in a relatively easy way. By virtue of a \texttt{Python} application, {\it i.e.}~a \texttt{Python} code, one is able to serve at the same time the front- and back-end, with a built-in internal web server and an external one, thanks to the {\tt Werkzeug} toolkit\footnote{\href{https://werkzeug.palletsprojects.com}{https://werkzeug.palletsprojects.com}}. The rendering of the website is done through different \texttt{HTML} templates, and it is handled by \texttt{Jin\-ja}\footnote{\href{https://jinja.palletsprojects.com}{https://jinja.palletsprojects.com}}, a web template engine. For each page one has to define the desired functions, that dictate the web application actions, into the {\tt Python} application, and assign the corresponding {\tt HTML} template.

\begin{figure*}[htbp]
\centering
\includegraphics[width=160mm,keepaspectratio]{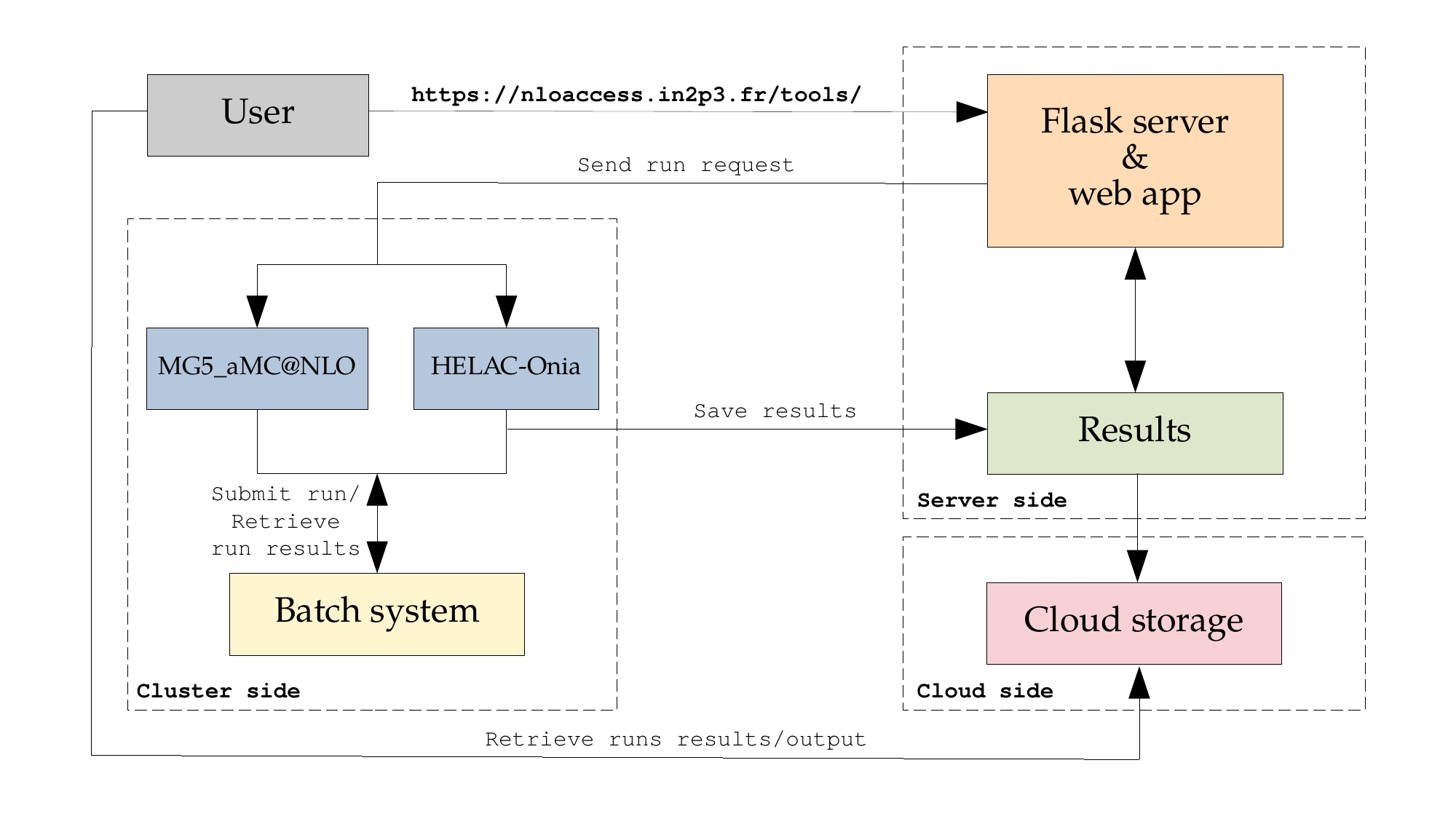}
\caption{Deployment layout of {\tt NLOAccess}. The three sides, together with the allowed interactions between an external user and the \NLOAccess platform are illustrated.}
\label{fig:deployment-diagram}
\end{figure*}

To describe the \NLOAccess design, we can divide the structure into three pieces: (a) the {\it server side}, (b) the {\it cluster side} and (c) the {\it cloud side}. A diagram illustrating this design and the interactions between these three parts is depicted on Figure~\ref{fig:deployment-diagram}.
The three sides serve the following functions:
\begin{enumerate}[(a)]
    \item the {\it server side} consists in the \Flask server and the internal \NLOAccess servers, where the code and the results are stored;
    \item the {\it cluster side} refers to the internal batch system, managed through {\tt HTCondor}\footnote{\href{https://htcondor.org}{https://htcondor.org}}. The codes present on \NLOAccess are able to communicate with the batch system, submitting runs on the cluster, retrieving and storing them in the \NLOAccess servers;
    \item the {\it cloud side} is represented by an {\tt OwnCloud}\footnote{\href{https://owncloud.com}{https://owncloud.com}} server, where the user personal folder, which contains the results for each of the requested runs, is securely stored and synchronised.
\end{enumerate}

As one can see from Figure~\ref{fig:deployment-diagram}, an external user is allowed to interact only with the server and the cloud sides through a web browser. On the server side, the user sees and interacts only with the front-end, served by the Flask application through the \NLOAccess website, on which the user can send run requests and wait for the results. The users are hence notified via e-mail when the run is finished. The results for each run are then stored in the personal user folder on the \texttt{OwnCloud} server. Such a folder is protected via the same password the user has chosen during the registration procedure needed to use the codes present on {\tt NLOAccess}.

Internally, on the cluster side, we notice that both \HELACOnia and \MG are able to interact with the batch system. For each requested run, they create their {\tt HT\-Con\-dor} job, submit it to the batch system, wait for the runs to be finished and retrieve the corresponding results. \NLOAccess is then able to retrieve the output and storing it in the corresponding cloud user folder. As already mentioned, the user is then also notified via e-mail for the run completion.

Since \NLOAccess allows every user to ask for a run request at any time, there is the need of keeping track of the different runs, associate them with the corresponding user, and put the results in their corresponding user folder. It is then natural to store some data about the user and the runs. To do so, we employ a {\tt PostgreSQL}\footnote{\href{https://www.postgresql.org/}{https://www.postgresql.org/}} database, that is ideal to be integrated in a {\tt Python} environment. The database securely stores data about the users\footnote{Data security is warranted. Password are securely encrypted and the user can ask for the cancellation of their data as the EU GDPR requires.}, the runs and the different processes that are generated (the latter information is kept for practical and statistical purposes).

Let us finally notice that this design allows any code that can be compiled and launched on a Unix shell to be included on the web platform.

\section{Features, usage and results}
\label{sec:results}

We turn now to the description of the usage of {\tt NLOAccess}. Generally speaking, the run requests are done via some dedicated forms or, for more articulated runs, via file input. The input file can either be built locally by the user and uploaded through the web pages, or it can be directly written down and uploaded via dedicated forms on the web platform. We stress that the building of the input file is the sole effort requested to the users, to which is offered all the computational power needed to perform their runs. The running time is a priori not limited, but a time limit can be set by the server.

As mentioned in Section~\ref{sec:intro}, at the moment, \NLOAccess offers the possibility to use two tools: {\tt MadGraph5\_aMC@NLO} and {\tt HE\-LAC\--Onia}. Although for each tool the commands needed to trigger a run are similar, we prefer to separately describe the usage of both codes on {\tt NLOAccess}. Indeed, these two codes, although sharing some aspects ({\it e.g.}~they both generate events, they involve Monte Carlo --MC-- integration, they are interfaced with external libraries etc.), are of different nature. While \HELACOnia is really a {\it code}, \MG is a {\it meta-code}, {\it i.e.}~a code generating other codes. As such, there are some small but substantial differences in the way the two codes are run, that reflect in some aspects, for instance the code re-usage. Since \HELACOnia is pre-compiled, it has a single executable associated to itself. Therefore, one has to run again the same process if some parameters such as PDFs, scale choices, or kinematical cuts, have to be changed. On the other hand, \MG allows one to generate a specific code for the requested process and compiles it. This allows the user to run the {\it same} code multiple times. Hence, while for \HELACOnia there can be a 1-to-$n$ correspondence between a process and the number of output folders, for \MG there is a 1-to-1 correspondence. Thus, for each run, the results will be stored in a sub-folder into the main process folder. These infrastructure differences, sum\-ma\-rised in Table~\ref{tab:summary-diff-HO-MG5}, reflect in the different usage of the two tools on {\tt NLOAccess}, as explained below in the following subsections.

\begin{table}[t]
\caption{Summary of infrastructure differences between \HELACOnia and {\tt MG5\_aMC@NLO}.}
\label{tab:summary-diff-HO-MG5}
\centering
\begin{tabular}{ccc}
\hline\noalign{\smallskip}
$ $ & \HELACOnia & \MG \\
\noalign{\smallskip}\hline\noalign{\smallskip}
code compilation & once & \parbox[t]{.28\columnwidth}{once for every {\tt generate/output} or {\tt launch} commands\\} \\
\hline\noalign{\smallskip}
running & \parbox[t]{.28\columnwidth}{run single executable for each run} & \parbox[t]{.28\columnwidth}{re-run the generated code for the requested process\\} \\
\hline\noalign{\smallskip} 
code re-usage & {\bf \large \ding{55}} & {\bf \large \ding{51}} \\
\noalign{\smallskip}\hline                                                            
\end{tabular}                                                                         
\end{table}

\subsection{{\tt HELAC-Onia}}
\label{sec:HO-usage}

Let us first start by illustrating how \HELACOnia can be used on {\tt NLOAccess}. \HELACOnia offers the generation of matrix elements and events for heavy quarkonium physics within the SM at Born order. It is based on Non-Relativistic QCD (NRQCD)~\cite{Bodwin:1994jh} and uses the off-shell recursion relations~\cite{Berends:1987me} to speed up the the matrix-element computation. Such a code is available at \href{https://nloaccess.in2p3.fr/HO/}{https://nloaccess.in2p3.fr/HO/}. 

Generally speaking, in the NRQCD framework, the cross section for a process (for instance, inclusive quarkonium production in $pp$ collisions) is written in a factorised form as follows:
\begin{equation}\label{eq:pp2psiX}
\begin{aligned}
    \sigma(pp \to {\Q} + X) & = \sum_{i,j,n} \int dx_1 dx_2 f_{i/p}(x_1)f_{j/p}(x_2)\\
    &\times \hat{\sigma}(ij \to Q\bar{Q}[n] + X) \langle \mathcal{O}_n^{\Q}\rangle
\end{aligned}
\end{equation}
where $f_{i/p}(x_1)$, $f_{j/p}(x_2)$ are the PDFs for the partons $i$, $j$ carrying a longitudinal momentum fraction $x_1$, $x_2$ from the two protons in the inital state, $\hat{\sigma}(ij \to Q\bar{Q}[n] + X)$ is the partonic cross section for producing a heavy-quark pair in the Fock state $n$, with $n = ^{2S+1}\!L_J^c$ ($S$ being the spin, $L$ the orbital and $J$ the total angular momentum of the heavy-quark pair, and $c=1,8$ indicating if the pair is in a colour singlet -- CS -- or colour octet -- CO -- state), and $\langle \mathcal{O}_n^{\Q}\rangle$ is the long distance matrix element (LDME) for the specific Fock state $n$, describing the transition between the heavy-quark pair and the observed quarkonium in the final state. 

\HELACOnia allows to calculate the convolution of these different terms for all such a kind of processes as in Eq.~(\ref{eq:pp2psiX}), not only for single but also for multiple quarkonia in the final state. It operates for different kinds of reactions (hadro-production, electro-production, photo-production etc.). This code is interfaced with {\tt LHAPDF}, allowing also for the computation of cross sections for collisions involving heavy ions by using nuclear PDFs (nPDFs). The code is also interfaced with parton shower generators like {\tt PYTHIA}. It also allows one to generate Les Houches Events (LHE, with the corresponding extension {\tt .lhe})~\cite{Alwall:2006yp}. 

In general, to compute a cross section with \HELACOnia on \NLOAccess, it is sufficient to write and submit an input file in the following general form:

\begin{center}
{\tt generate $\{$ process $\}$} \\
{\tt set $\{$ parameter $\}$ = $\{$ value $\}$} \\
{\tt $\vdots$} \\
{\tt decay $\{$ particle $\}$ @ $\{$ BR $\}$ } \\
{\tt $\vdots$} \\
{\tt launch} \\
\end{center}

One can highlight four different kinds of commands:
\begin{enumerate}[(i)]
    \item the {\tt generate} command is needed to indicate the desired process to be generated. For instance, if one wishes to generate our example in Eq.~(\ref{eq:pp2psiX}) for a $J/\psi$ in CS, it is sufficient to write ``{\tt generate p p > cc$\sim$(3S11) j}", where the corresponding syntax for the $J/\psi$ in the CS state and for a jet (representing the set of partons created in the hard scattering which is denoted as $X$ in Eq.~(\ref{eq:pp2psiX})) is used. The syntax for the particles and the quarkonium states can be found on {\tt NLOAccess}\footnote{See \href{https://nloaccess.in2p3.fr/tools/sm-particles}{https://nloaccess.in2p3.fr/tools/sm-particles} and \href{https://nloaccess.in2p3.fr/tools/quarkonia}{https://nloaccess.in2p3.fr/tools/quarkonia} for SM particles and quarkonia respectively.};
    \item the {\tt set} command allows one to change every parameter defined within {\tt HELAC-Onia}. Through this command, one has control on several classes of parameters: 
    \begin{itemize}
    \item collision parameters, {\it i.e.}~which particles are in the initial state;
    \item theory parameters such as the factorisation and renormalisation scales or the value of the coupling constants;
    \item PDF related parameters ({\it e.g.}~whether to use {\tt LHAPDF} and which (n)PDF set to use);
    \item MC setup parameters;
    \item kinematical cuts;
    \item quarkonium specific parameters ({\it e.g.}~the value of the LDMEs);
    \item physical constants (like the quark, bosons and lepton masses); 
    \item parameters specifying the desired kind of output ({\tt ROOT}, {\tt Gnuplot}, {\tt TopDrawer} and/or {\tt .lhe} files). 
    \end{itemize}
    
    A list of {\tt set} commands is then required if the user wishes to use different parameters for a run\footnote{A dedicated user guide for the \HELACOnia syntax is available at \href{https://nloaccess.in2p3.fr/tools/downloads/HO_online_guide_v01.pdf}{https://nloaccess.in2p3.fr/tools/downloads/HO\_online\_guide\_v01.pdf}.};
    
    \item the {\tt decay} command allows one to define a specific BR for the decay of a particle. For instance, if one wants a $J/\psi$ to decay in di-muons with a BR of 6\%, is it sufficient to use {\tt decay cc$\sim$(3S11) > m+ m- @ 0.06}; the corresponding computed cross section will then be multiplied by the desired BR;
    \item the {\tt launch} command triggers the computation.
\end{enumerate}
The syntax for the particles and the naming of the most important variables are illustrated on the online guide for \HELACOnia which we have developed and made available on {\tt NLO\-Ac\-cess}.

An input file can be directly uploaded or built on the web platform using the dedicated forms. This will trigger the calculation for the desired process and parameters and, if the calculation will be successful, the user will receive an e-mail containing the link of their personal {\tt OwnCloud} folder, into which they will find the plots done for them on the fly. This relatively simple and automated procedure allows the \NLOAccess users to obtain results with \HELACOnia in a straightforward way. 

A last comment about the usage of \HELACOnia is in order. As mentioned at the beginning of Section~\ref{sec:results} and also summarised in Table~\ref{tab:summary-diff-HO-MG5}, this tool provides a single executable that is used for every run. Linked to this executable, there is an analysis code, that is used to create and fill the histograms given as output of a calculation. At the moment, we are then limited to offer to the users the generation of the plots for the total cross section, and the single differential cross sections in terms of rapidity $y$ and transverse momentum $P_T$ of each particle in the final state. If a user is interested to get more complex kinematical distributions, they can be obtained through the analysis of the LHE event file.

\subsection{{\tt MG5\_aMC@NLO}}
\label{sec:MG5-usage}

We now turn to the \MG usage on {\tt NLOAccess}. First, we emphasise that the availability of \MG on \NLOAccess represents {\it for the first time} the opportunity to use such a code on a web interface in its full version, {\it i.e.}~allowing also for NLO computations. Indeed, in the past, an online platform for \MG existed, with a couple of mirrors around the world, but the allowed calculations were limited to leading order. \MG on \NLOAccess is reachable at 
\href{https://nloaccess.in2p3.fr/MG5/}{https://nloaccess.in2p3.fr/MG5/}. 

Although such a code is very well known and widely used, especially for the LHC physics, let us remind that \MG is a framework for SM and BSM phenomenology that provides all the necessary elements to compute cross sections, generating pertubative short-distance events and matching them with parton-shower programs. The processes can be simulated at LO accuracy for any user-defined Lagrangian, and at NLO for a subset of models. We notice here that the current version of \MG already allows the computation of cross-sections and the generation of events at NLO in collinear QCD for symmetric nuclear beam ($AA$) collisions, as this requires only to indicate an \texttt{LHAPDF} ID for nPDFs in the run card. The extension to asymmetric hadronic collisions (\textit{e.g.}~$pA$ collisions), that will allow the automated computation of nuclear modification factors at fixed-order in QCD, is now being finalised. Preliminary results are presented in Ref.~\cite{Safronov:2022uuy}. We guide the reader to  Ref.~\cite{Alwall:2014hca} and to their Launchpad website\footnote{\href{https://launchpad.net/mg5amcnlo}{https://launchpad.net/mg5amcnlo}} for all the information about this framework.

Moving on, we first recall that \MG is a meta-code. This allows one to separate the usage of such a tool in two stages: the code generation and the code running. This clearly reflects in the usage of \MG on the web platform. The first stage consists in the generation of the code via a simple form, that requests three values: the name of the model, the process to be generated, the name of the corresponding output folder. Such a first stage can be also obtained uploading a simple input file in the following form:
\begin{center}
{\tt import model $\{$ model\_name $\}$} \\
{\tt generate $\{$ process $\}$} \\
{\tt output $\{$ folder\_name $\}$} \\
\end{center}
where again the curly brackets indicates specific values for the three parameters.
By default, the SM is used as a model. If an NLO computation is requested using the corresponding \MG syntax, the {\tt loop\_sm} model will be then imported. Notice that, although both \HELACOnia and \MG works for $2 \to n$ processes, in the latter tool no hadron can be generated with a fixed-order (FO) computation, but only through parton shower generators. Moreover, we underline that, at present, the sole way to compute quarkonium-production cross sections at FO in \MG is by applying the Colour-Evaporation Model~\cite{Fritzsch:1977ay,Halzen:1977rs} (CEM). This approach relies on the principle of quark-hadron duality and, to be applied, it requires dedicated cuts on the invariant mass of the heavy-quark pair. These cuts can be implemented in an analysis code for a specific process generated with \MG involving an heavy-quark pair in the final state\footnote{As for the moment the upload of parts of code is not allowed on {\tt NLOAccess}, we invite the users to contact the NLOAccess Team if different analysis  files from the ones already available are needed.}. An example of such a calculation can be found in Ref.~\cite{Lansberg:2020rft}, where \MG was used to apply the CEM at NLO for single and double quarkonium hadro-production at Tevatron and the LHC. 
\begin{figure}[ht]
\centering
\includegraphics[width=\columnwidth,keepaspectratio]{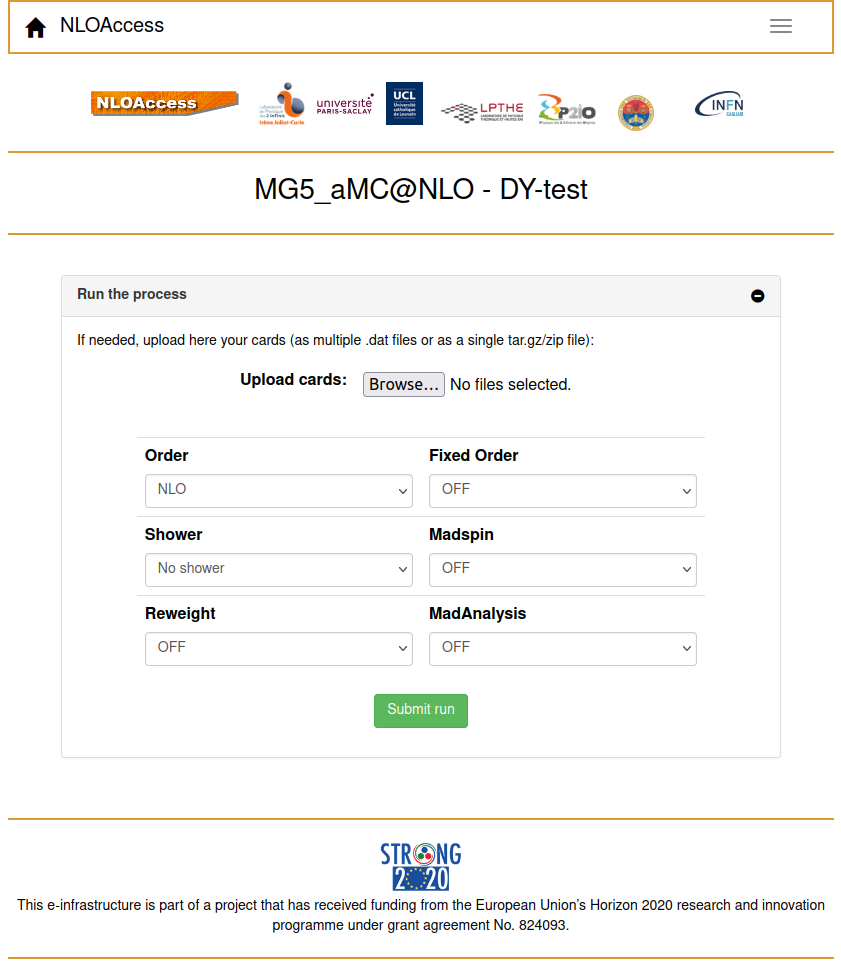}
\caption{Example of the form on \NLOAccess to run the NLO \MG code.}
\label{fig:MG5-form}
\end{figure}

The second stage to use \MG on \NLOAccess is the code running. The users can find their generated codes on their personal \MG code database on {\tt NLOAccess}. Once they select the desired code to be run, the user finds a form with few drop-down menus and a form where to upload their cards. The form changes whether the code was generated using the LO or NLO syntax of {\tt MG5}. An example of a form for an NLO code is shown in Fig.~\ref{fig:MG5-form}. Both kind of codes can be run either in the FO mode, or can be interfaced with a parton shower. At the moment, within {\tt NLOAccess}, two parton-shower generators are interfaced with {\tt MG5}: {\tt PYTHIA} (both version {\tt 6} and {\tt 8}) and {\tt HERWIG 6}~\cite{Corcella:2000bw}. Other tools can be used for a run, for instance {\tt MadAnalysis5}~\cite{Conte:2012fm} (for plotting and analysis purposes), {\tt MadSpin}~\cite{Artoisenet:2012st} (for particle decays) or to apply some reweighting procedures~\cite{Mattelaer:2016gcx,Kalogeropoulos:2018cke}. Again, once a run is performed and completed, the user is alerted via e-mail and the results are uploaded and securely stored in the user personal {\tt OwnCloud} folder. Contrary to {\tt HELAC-Onia}, for which the generation of {\tt .lhe} files has to be requested, \MG automatically provides the events under the LHE format for all the runs but for the FO NLO ones. In the latter case, plots for total and differential cross-sections are automatically generated. 
Once again, the user can obtain the desired results, with a couple of simple steps, with this automated two-step procedure.

\subsection{Further general features}

Having described the usage of both tools accessible via {\tt NLO\-Ac\-cess}, we now conclude by describing some further general features of the web platform. 

To enhance the user-friendliness of the website, we allow the users for the management of their runs. They are in fact able to check their live run status on a dedicated page, and to remove a run - or multiple runs - if needed. For each run, the number of sub-jobs running, in idle state or completed is also shown. The information on every run is then retrieved in the personal run history of the user. In this way, the user will retrieve information on the tool that was used, the generated process, the date and time of the running and the duration of the run itself. The total amount of time taken by all the runs performed for the user is also given.

\section{Summary and outlook}
\label{sec:conclusions}

In this paper, we presented {\tt NLOAccess}, an online platform devoted to automated computation of physical observables for collider physics, part of the EU Horizon 2020 integrated initiative ``STRONG-2020". At pres\-ent, \NLOAccess allows any kind of user, from students and non-experts to people with experience, to generate events and compute the cross section, in the collinear factorisation approach, for several kinds of processes and for different initial and final state particles involving also heavy-ion collisions and quarkonium production. Noticeably, \NLOAccess also allows for the first time to run on the web \MG in its full NLO version. We showed how, with few steps, any user is able to generate and run codes, obtaining automatically plots done on the fly, securely stored in a personal cloud folder. Such an online platform allowed so far around 400 users from all over the world to perform over 4000 runs.  Thanks to its design, \NLOAccess allows in principle any code that can be compiled and launched on a Unix shell to be included, hence it represents an opportunity for who is interested in promoting their own scientific codes.

Within the \NLOAccess project, the team is also involved in code developments. We mention here the extension of \MG to the case of asymmetric hadronic collisions, {\it i.e.}~collisions between two hadronic beams of different nature ({\it e.g.} proton-nucleus, nucleus $A$-nucleus $B$, pion-proton). Preliminary results are presented in Ref.~\cite{Safronov:2022uuy}, and will be made available soon on {\tt NLOAccess}. These results will allow one to automatically calculate at fixed-order the nuclear modification factors, that describes the nuclear effects affecting the PDFs of quarks and gluons inside a bound nucleon. In parallel to this development, there is also ongoing effort for extension and validation of lepton-hadron collisions in {\tt MG5}. Finally, we also mention the creation of a {\tt Python3} compliant version of {\tt HELAC-Onia}, that is available under request.

Given the experience maturated since the beginning of the project, where \NLOAccess was used not only as a service for the community, but also to train students and non-expert during several tutorials and hands-on session, we believe that such a kind of project successfully shows all the benefits of online access to such  automated codes for everyone.

\begin{acknowledgements}
We thank L.-A.~Couturi\'e, V.~Lafage, J.-P.~Lansberg, O.~Mattelaer and H.-S.~Shao for their help and suggestions at different stages of the developement of the \NLOAccess platform. We also thank O.~Mattelaer for a careful reading of the manuscript.
This project is supported by the European Union's Horizon 2020 research and innovation programme under Grant agreement no.~824093.
This project has also received funding from the Agence Nationale de la Recherche (ANR) via the grant ANR-20-CE31-0015 (``PrecisOnium'') and via the IDEX Paris-Saclay "Investissements d’Avenir" (ANR-11-IDEX-0003-01) through the GLUODYNAMICS project funded by the "P2IO LabEx (ANR-10-LABX-0038)".
This work  was also partly supported by the French CNRS via the IN2P3 project GLUE@NLO, via the Franco-Polish agreement COPIN-IN2P3 and via the Franco-Polish EIA (GlueGraph).

\end{acknowledgements}

\bibliographystyle{utphys}       % APS-like style for physics
% \bibliography{nloaccess}   % name your BibTeX data base

\providecommand{\href}[2]{#2}\begingroup\raggedright\endgroup

\end{document}